\documentclass{article}
\usepackage{amsmath}
\usepackage{amsfonts}
\usepackage{amssymb}
\usepackage{epsfig}
\usepackage{wrapfig}
\usepackage{siunitx}
\usepackage{authblk}
\usepackage[doublespacing]{setspace}
\usepackage[textsize=tiny]{todonotes}
\usepackage{listings}
\usepackage[numbers]{natbib}
\usepackage{hyperref}

\usepackage{color}

\newcommand{\beq}{\begin{equation}}
\newcommand{\eeq}{\end{equation}}

\newcommand{\ben}{\begin{eqnarray}}
\newcommand{\een}{\end{eqnarray}}

\newcommand{\Pseudo}{\textit{Pseudomonas aeruginosa}}
\newcommand{\pseudo}{\textit{P. aeruginosa}}

\begin{document}

\title{Computational tools for the multiscale analysis of Hi-C data in bacterial chromosomes}

\author{Nelle Varoquaux$^{1,2}$, Virginia S. Lioy$^{3}$, Fr\'ed\'eric Boccard$^{3}$, Ivan Junier$^{1,2*}$}
\affil{$^{1}$ CNRS, TIMC-IMAG, F-38000 Grenoble, France\\$^{2}$ Univ. Grenoble Alpes, TIMC-IMAG, F-38000 Grenoble, France\\$^{3}$ Université Paris-Saclay, CEA, CNRS, Institute for Integrative Biology of the Cell (I2BC), 91198, Gif-sur-Yvette, France}

\date{}
\maketitle

$^{*}$Corresponding author: \texttt{ivan.junier@univ-grenoble-alpes.fr}
\maketitle
\newpage

\noindent{\textbf{Running head}: Multiscale analysis of Hi-C data in bacterial chromosomes}\\

\section*{Summary}
Just as in eukaryotes, high-throughput chromosome conformation capture (Hi-C)
data have revealed nested organizations of bacterial chromosomes into
overlapping interaction domains. In this chapter, we present a multiscale
analysis framework aiming at capturing and
quantifying these properties. These include both standard tools (e.g.~contact
laws) and novel ones such as an index that allows identifying loci involved in
domain formation independently of the structuring scale at play. Our objective
is two-fold. On the one hand, we aim at providing a full, understandable
Python/Jupyter-based code which can be used by both computer scientists as well as 
biologists with no advanced computational background. On the other hand, we
discuss statistical issues inherent to Hi-C data analysis, focusing more particularly
on how to properly assess the statistical significance of results.
As a pedagogical example, we analyze data
produced in \Pseudo, a model pathogenetic bacterium. All files (codes and
input data) can be found on a github repository. We have also embedded the
files into a Binder package so that the full analysis can be run on any
machine through internet.

\section*{Key words}
Multiscale analysis of Hi-C Data; Chromosome Interaction Domains; Nested organization; Python/Jupyter tools

\section{Introduction}
Functional organization of bacterial genomes spans multiple scales, from a few
base pairs to the entire genome~\cite{Junier:2014cpa,Touchon:2016fp}. From the viewpoint of
chromosome structuring, supercoiling domains ($\SI{10}{kb}$
scale)~\cite{Postow:2004fp} and macrodomains (Mb scale)~\cite{Valens:2004um,Lioy:2018ca} are
for instance known to be functionally
important~\cite{Lagomarsino:2015kg,Junier:2018cl}. In the last five years or so, high-throughput chromosome conformation capture (Hi-C) data have further
revealed the existence of nested sets of overlapping chromosome interaction
domains (CIDs) that span a wide range of scales, from tens to hundreds
kb~\cite{Le:2013ci,Le:2014df,Wang:2015ee}. Just as in eukaryotes, important
questions concern the functional relevancy of these domains
as well as physical mechanisms underlying their formation. To tackle these problems, various tools have been developed, mostly in the context of eukaryotes (see~\cite{Haddad:2017ea} and references therein) to quantify the nested organization of chromosomes. 

Here, we present a multiscale analysis framework we have developed to tackle
the functional structuring of \Pseudo's chromosome~[in prep]. Due to the
generality of our tools/codes, we aim at making them freely available. They should be useful, in particular, to biologists interested in manipulating, analyzing, visualizing Hi-C data, helping them to rationalize the multiscale domain organization of chromosomes, be it in bacteria or in eukaryotes. Our approach is based on an original 2D pattern analysis of Hi-C heat maps. In particular, we introduce an index~(the {\it frontier index}) that quantifies the implication of any chromosomal locus to domain frontiers that may occur at any scale.

\section{Methods}
The Methods section falls into the following subsections:
\begin{enumerate}
	\item We explain how to use our framework, providing two possibilities: by downloading our code and data or by performing the analysis {\it via} a freely available cloud server.
	\item We introduce the data and organism we have been analyzing.
	\item We discuss technical aspects of our computational approach, which relies heavily on the
use of matrices.
	\item We show how to compute contact laws, i.e.\ average contact frequencies between pairs of loci as a function of their genomic distance.
	\item We propose a simple method using matrix
derivatives to highlight the frontiers associated with the interaction domains
of Hi-C heat maps.
	\item We propose a new index, the {\it frontier index},
that quantifies the implication of any locus into any (i.e.~at any scale) such
frontier.
	\item We discuss randomization procedures to assess the statistical significance of the frontier index.
\end{enumerate}

\subsection{Using our framework}
While the approach described here can be implemented in many different
languages, we provide a Python implementation from scratch using the basic
tools of the Scientific Python ecosystem. Why the Scientific Python stack?
Because the core libraries of the Scientific Python ecosystem, \texttt{numpy}
and \texttt{scipy}, contain a wide set of functions for processing scientific
data easily and efficiently. While we here focus on the bacterium \textit{P. aeruginosa},
our code can scale to eukaryotes, including human genomes, with minor modifications. In
addition, the Python ecosystem has grown dramatically over the years, in a
broad spectrum of disciplines, ranging from astronomy to climate science and
biology.

As a supplementary material to this chapter, we provide a Jupyter notebook (\texttt{Methods.ipynb})
containing the code to reproduce all of the figures presented here as well as
details on specific points of the methods. In the text, we use the notation ``notebook X''  to refer to any section X of the \texttt{Methods.ipynb} notebook. 
In addition, those notebooks can be executed for free online thanks to mybinder
\url{https://mybinder.org/v2/gh/TREE-TIMC/2020-multi-hic/master?filepath=index.ipynb}. For those who would rather like to reproduce the results locally, the
supplementary materials are hosted in a GitHub repository:
\url{https://github.com/TREE-TIMC/2020-multi-hic}. The repository contains
detailed instructions to install all the dependencies and execute the code.

\subsection{Input data}
Here, we deal with Hi-C data obtained in a \pseudo\ strain lacking the condensin MksBEF. Data were collected during exponential growth in presence of glucose~[in prep]. Raw data were normalized according to
the ICE procedure~\cite{Imakaev:2012dd} using the Python library
\texttt{iced}~\cite{Varoquaux:iced}: ICE supposes equal visibility for
all chromosomal loci, meaning that the cumulative contact frequency of every
locus with all other loci is almost identical (notebooks 2.1). To that
matter, Hi-C data were produced from fastq files using the HiC-Pro
software~\cite{Servant:2015iu}, which generates both raw and ICE-normalized
data with specific binning of the genomes (here we discuss
$\SI{10}{kb}$ resolution of the matrix). Interaction frequencies generated by
HiC-Pro are provided by a table made of three columns and as many rows as
there are pairs of chromosomal loci with a non-zero interaction frequency (see
\texttt{data/res10000.txt}): the two first columns indicate the pair of interacting
loci and the third row the Hi-C value. Here, for clarity, we further normalize
data such that the cumulative contact frequency of every locus is (almost)
equal to $1$ (notebook 2.2). We note here that a locus
corresponds to an indexed interval along the genome (it is thus often referred
to as a bin) and, hence, depends on the discretization, i.e.\ there are
respectively $6276$ and $628$ loci at $\SI{1}{kb}$ and $\SI{10}{kb}$
resolutions.

\subsection{Matrix-based approach using \texttt{numpy} arrays}
\begin{figure}
\center
\includegraphics[width=\linewidth]{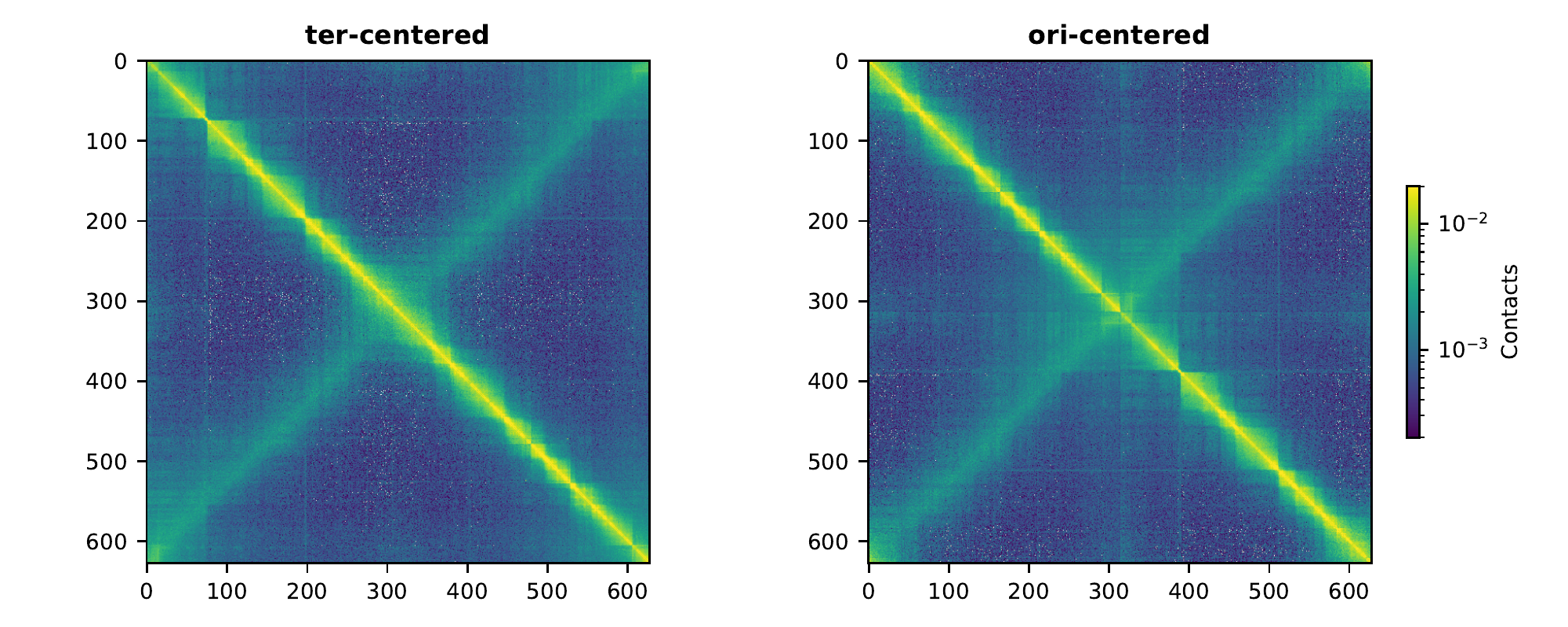}
\caption{
\label{fig:ori_ter}
{\it Different origins for the visualization of Hi-C matrices}. Left
panel: the heat map is centered on the terminus of replication (the upper left
origin corresponds to the origin of replication). Right panel: the heat map is
centered on the origin of replication (the upper left origin corresponds to
the terminus of replication). Note the presence of two diagonals. The most
pronounced one reflects the local organization of the chromosome. The second
one reflects the longitudinal organization of the
chromosome~\cite{Lagage:2016bm} and more specifically, the
``zipping/extruding'' action of SMC condensins~\cite{Burmann:2015be}.
}
\end{figure}

Our computational analysis, manipulation, and visualization heavily rely on
the use of matrices (Notes~\ref{matrices}). To that end, we use the array type
from \texttt{numpy} Python module. As an example of the
matrix power, changing the center of the Hi-C heat map (Fig.~\ref{fig:ori_ter})
only requires one line of code (notebook 2.3). Note that in the
following we distinguish {\it heat maps} akin to visualization (Notes~\ref{colormap}) from
{\it matrices} akin to mathematical manipulations. Note also that bin
enumeration starts at $1$ in HiC-Pro contact frequency tables, while
(\texttt{numpy}) matrix indexing starts at $0$.

\subsection{Estimating the contact law $P(s)$}
\begin{figure}
\center\includegraphics{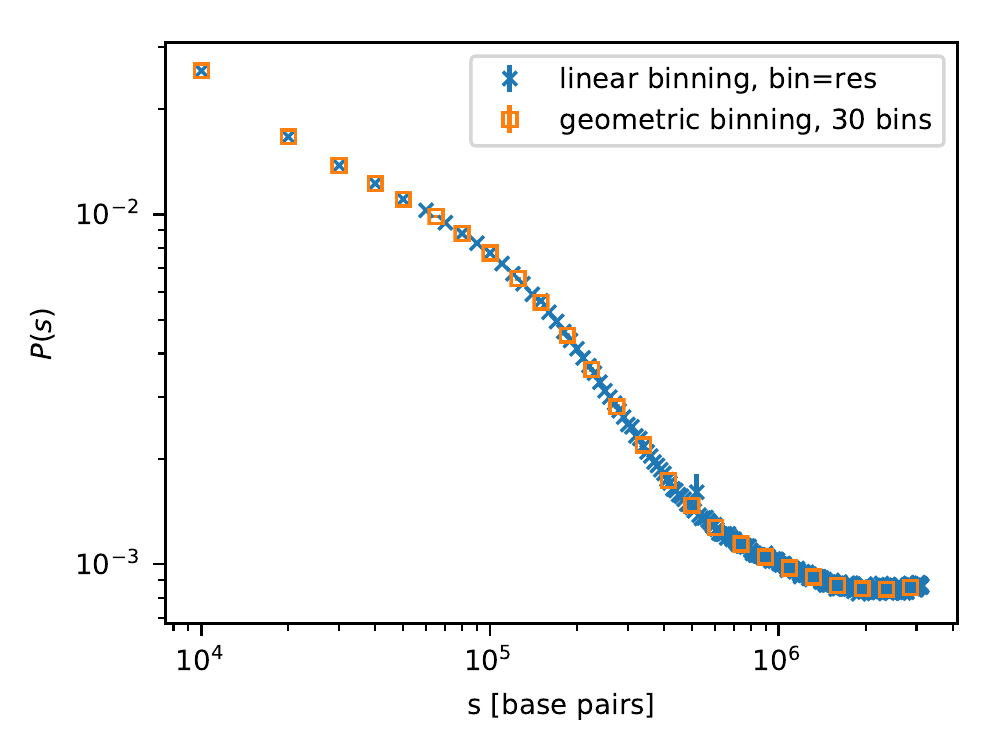}
\caption{
\label{fig:Ps} 
{\it Contact law $P(s)$}. Error bars, which are smaller than the symbols except for the blue point at $s\simeq \SI{500}{kb}$, correspond to the standard error of the mean. As expected, a proper geometric binning (in orange) leads to the same contact law as a linear binning (in blue) but using a smaller number of bins (that are equally spaced in the logarithmic scale). The abnormal point for $s\simeq \SI{500}{kb}$ reflects a mapping problem~(Notes~\ref{outliers}). While abnormal contact frequencies can be corrected (notebook 3.2), their impact is greatly reduced using the geometric binning.
}
\end{figure}

In addition to its necessity to assess the relevancy of (polymer) physical
models of chromosome structuring~\cite{Mirny:2011cl}, contact laws provide a convenient summary of the multiple scales of chromosome structuring~(Notes~\ref{outliers}). In this regard, it
is common to use a log-log plot in order to visualize the full range of
genomic scales as well as the full range of interaction frequencies
(Fig.~\ref{fig:Ps}).  Note finally, that for a log-log plot of $P(s)$, it is
often  more appropriate to use a geometric (instead of linear) binning for the
genomic distance $s$. It reduces for instance mapping errors on the estimation of $P(s)$ (Fig.~\ref{fig:Ps}).

In practice, a contact law $P(s)$ together with its corresponding standard
deviation ($Std(s)$) and standard error of the mean ($Stde(s)$) are computed
following these steps (notebook 3.1):

\begin{enumerate}
\item {\it Binning of the genomic distances}. The binning
can be either linear or geometric. In the latter case, one needs to specify
the total number of bins ($N_b$) and the shortest distance ($s_0$). Then,
calling $L_g$ the genome length, the binning between $s_0$ and $L_g/2$ reads
$B=[s_0,\dots,s_0\times r^i,\dots,s_0\times r^{N_b}]$ with
$r=\left(\frac{Lg}{2s_0}\right)^{1/N_b}$.

\item {\it Extracting the diagonal indexes associated with genomic distances using \texttt{numpy.diag()}.} Diagonal numbers (Notes~\ref{diagonals}) for the $i^{th}$ bin of the genomic distances correspond to the integers in $[\text{int}(B[i]/\rho),\text{int}(B[i+1]/\rho)[$, where $\rho$ is the Hi-C resolution (10 kb here) and where $\text{int}(\bullet)$ stands for the integer part function.

\item {\it Computing statistics}. First, we compute the mean genomic distance, $s$, associated with each bin. Second, considering the values from the diagonals
extracted in stage $(2)$, we compute the mean and standard deviation of the
frequency of contacts, respectively leading to $P(s)$ and $Std(s)$. The standard
error of the mean is then given by $Stde(s)=Std(s)/\sqrt{N-1}$.

\end{enumerate}

\subsection{Highlighting the frontiers of a Hi-C heat map}
\begin{figure}[htp]
\center\includegraphics[width=\linewidth]{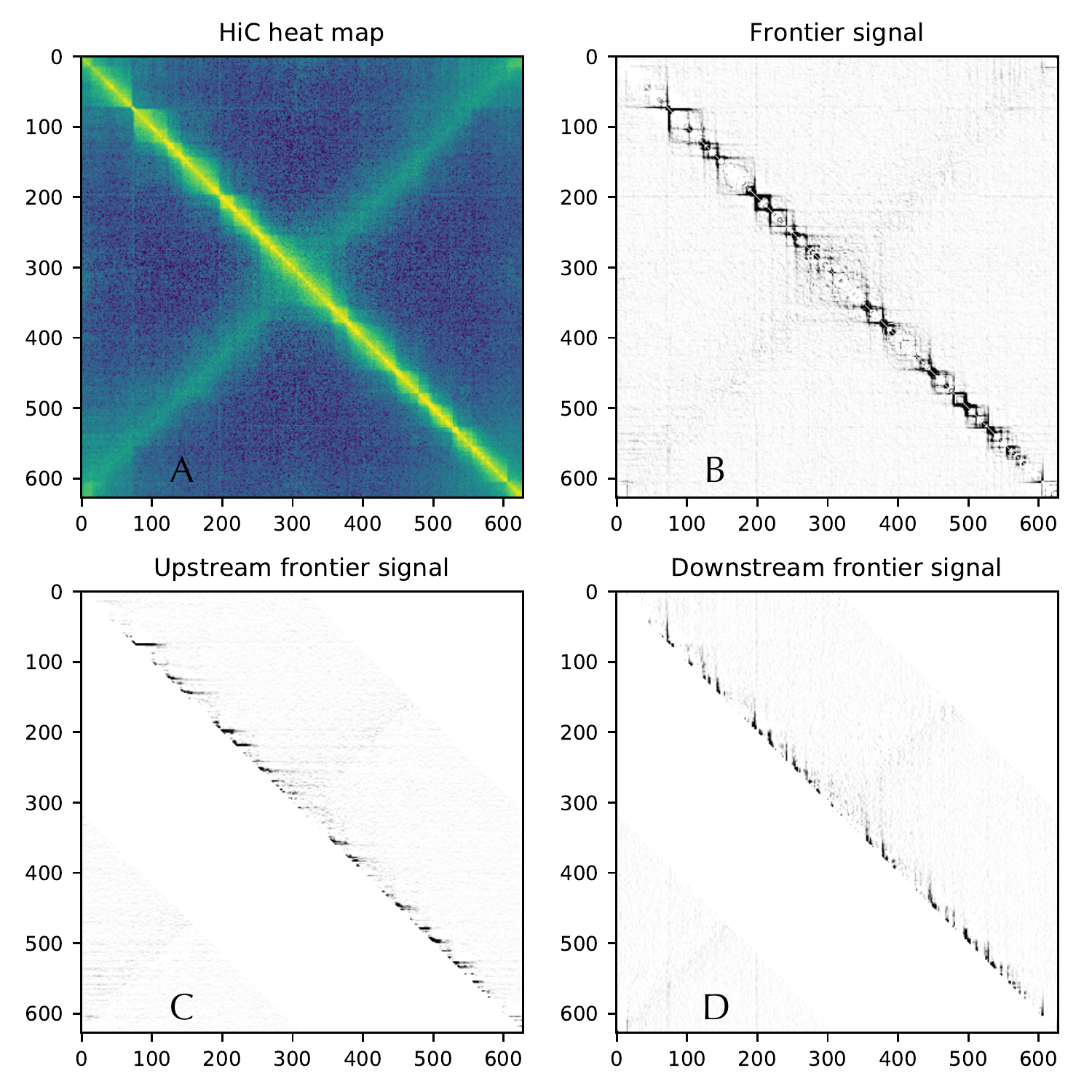}
\caption{
\label{fig:frontiers} 
{\it Frontiers signal associated with a Hi-C matrix}. (A) Original Hi-C
matrix. (B) Upstream frontier signal reflecting the implication of any pixel
to a upstream frontier of a CID (or of domains embedding multiple
CIDs~(Notes~\ref{frontiers})). (C) Same as B but for the downstream frontiers.
For (B) and (C), we actually show only half the matrix (the other half is
generated using matrix transposition). The reason is to stress two important
points. First, we recall that because of the chromosome circularity, genomic
distances (in matrix indexes) are smaller than $N/2$ (with $N$ the number of
loci). Second, as a consequence of chromosome circularity, again, the frontier
value associated with a pair $ij$ of loci such that $i<N/2$ and $j>N/2$, and  must be read from the entry $[j,i]$ of the gradient matrix (left lower conner part of the matrix) and not from the entry $[i,j]$ (right upper conner part of the matrix, left blank).(D)$=$(B)$+$(C) plus a matrix transposition (notebook 4).
}
\end{figure}

A chromosome interaction domain (CID)~\cite{Le:2013ci} in a Hi-C heat map (Fig.~\ref{fig:frontiers}A) is delineated by both
upstream (Fig.~\ref{fig:frontiers}B) and downstream (Fig.~\ref{fig:frontiers}C) frontiers, that is, by sharp decreases of contact frequencies between loci within the CID and loci outside the CID located upstream and downstream, respectively~(Notes~\ref{frontiers}).
 
Here, we provide a simple method based on the derivative (equivalently, gradient, Notes~\ref{gradient}) of the Hi-C matrix to quantify the implication of any pixel of the heat map in both upstream (Fig.~\ref{fig:frontiers}B) and downstream (Fig.~\ref{fig:frontiers}C) frontiers, the combination of the two signals leading to a matrix that highlight all the frontiers of the Hi-C heat map (Fig.~\ref{fig:frontiers}D). The associated code can be found in the function \texttt{generate\_frontiers()} of notebook 4. It consists in:

\begin{enumerate}
    \item Subtracting the contact law from the Hi-C matrix so that to focus on
    variations associated with the presence of domains.
    \item Smoothing the data using a Gaussian filter so that to be able to
    safely compute derivatives of the matrix.
    \item Computing the derivative of the Hi-C matrix.
    \item Building the frontier matrices by noting that the upstream (downstream) frontiers consist of the positive (negative) signal associated with the first (second) axis of the derivative, that is, with the derivative along the rows (columns). We thus set to 0 the value of the gradient whenever it is negative (positive) and, for the downstream frontier, we take the absolute value of the signal.
\end{enumerate}

\subsection{Frontier indexes}
\begin{figure}
\center\includegraphics[width=\linewidth]{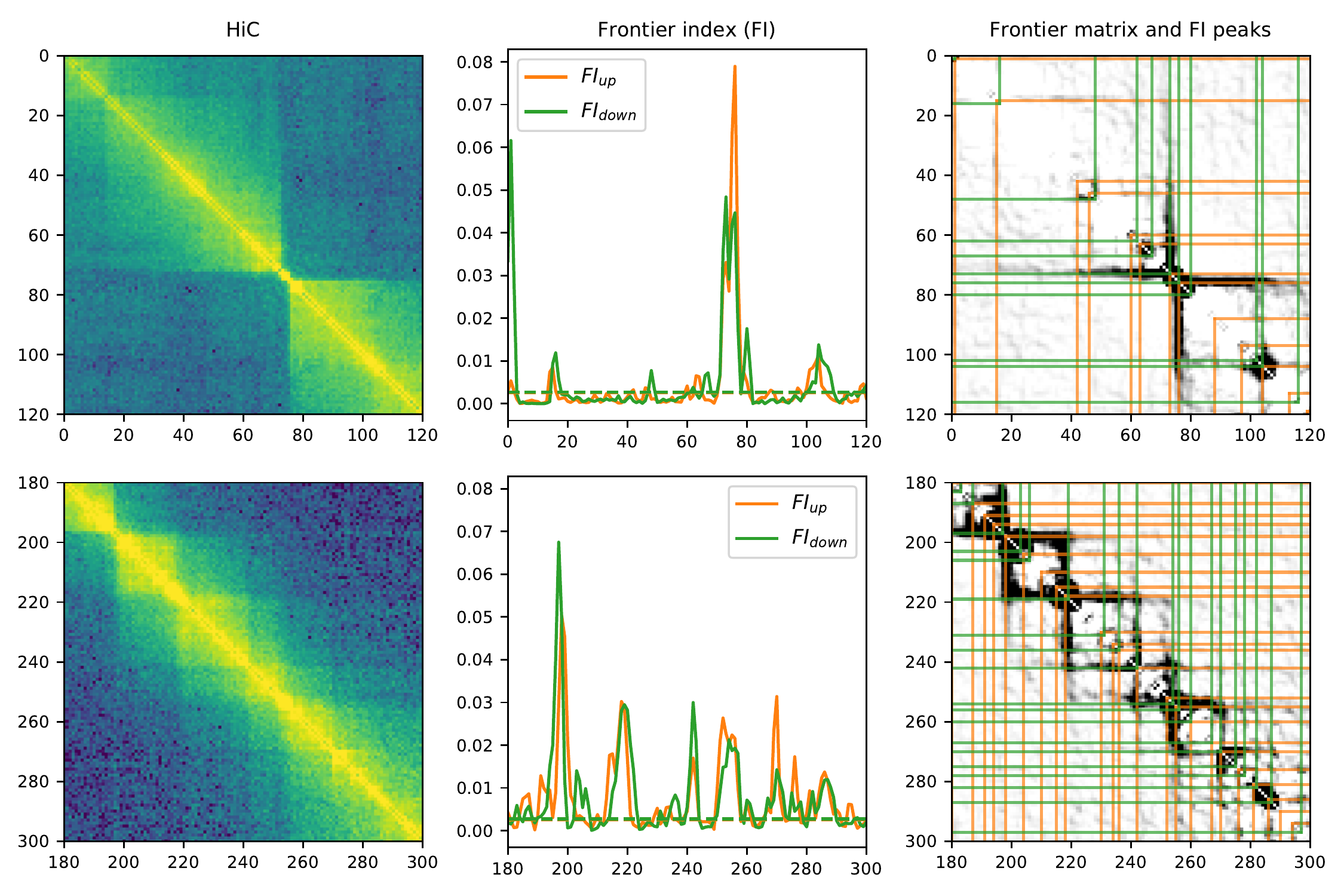}
\caption{
\label{fig:peaks} 
{\it Identification of frontiers}. For both rows: the left panel corresponds to a zoom in of the HIC data; the middle panel shows the corresponding values of the upstream and downstream frontier indexes; the right panel shows the frontier matrix together with downstream and upstream lines associated with the peaks identified for the upstream and downstream frontier index, respectively. To that end, we conserved the peaks that were above the median values of all the peaks (notebook 5).
}
\end{figure}

For both the upstream and downstream 2D frontier signals, we respectively define a upstream ($FI_{up}(i)$) and downstream ($FI_{down}(i)$) frontier index of a locus $i$ as the cumulative signal of the corresponding frontier matrix in the downstream, respectively upstream direction. To mitigate noise sources, we further limit the signal to loci that are below a certain distance $s_{max}$ of $i$ -- here, we take $s_{max}=\SI{500}{kb}$. More precisely, if $F_{up}$ and $F_{down}$ are the frontier matrices, then the frontier indexes $FI_{up}$ and $FI_{down}$ of a locus $i$ are given by (notebook 5):

\ben
FI_{up}[i] &=& \sum_{k>0,s_{i,(i+k)\%N}\leq s_{max}}^{k=N/2}F_{up}[i,(i+k)\%N]\\
FI_{down}[i] &=& \sum_{k>0, s_{i,(i-k+N)\%N}\leq s_{max}}^{k=N/2}F_{down}[(i-k+N)\%N,i]
\een

\noindent
where $\%$ indicates the modulo operator and where $s_{i,j}$ indicates the genomic distance between loci $i$ and $j$. Note, here, that we use the Python matrix notation to explicit the implementation -- the mathematical definition is lighter as, for instance, we would have $FI_{up}(i)=\sum_{j,s_{i,j}\leq s_{max}}F_{up}(i,j)$. 

These definitions lead to profiles of frontier indexes from which we can identify peaks of local maxima (middle panels in Fig.~\ref{fig:peaks}). We can then filter the peaks such that the  value of the frontier index at the peak is larger than a threshold. This threshold can either be provided by the user or estimated in an {\em ad hoc} manner (see e.g. legend of Fig.~\ref{fig:peaks}), or it can be set by estimating a p-value for each candidate loci, as we explain below. Once these peaks have been identified, we can report them on the original matrix or on the frontier matrix to visually check that they indeed correspond to frontiers (Notes~\ref{fullmulti}).

\paragraph{Estimating a p-value} The subtlety here lies in building a proper
null model so that to define a relevant false discovery rate (FDR).  We propose a simple
randomization procedure akin to shuffling the contact counts associated with
the same genomic distance $s$. The statistic we use is the frontier index
$FI(i)$. As a result, we propose to randomize the frontier matrix $F_{ij}$
directly ($F$ can be either the upstream or downstream frontier matrix). This leads to the following procedure (notebook 6):

\begin{figure}
\center\includegraphics[width=\linewidth]{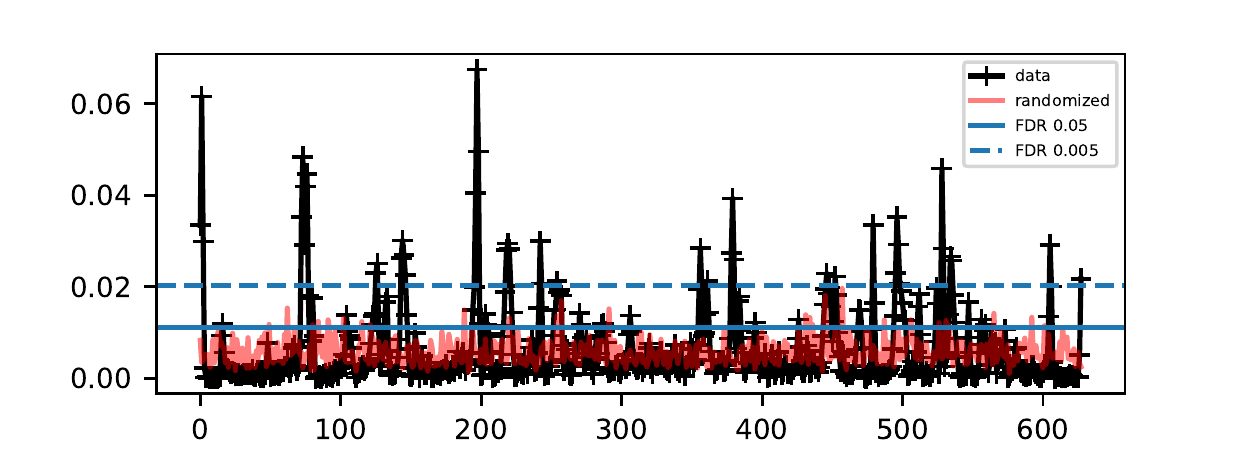}
\caption{
\label{fig:FI_FDR} 
{\it Assessing the statistical significance of the frontier index}. The black curve indicates the upstream frontier index for \pseudo. The red translucent curve corresponds to a randomization of the upstream frontier matrix, revealing an over-estimation of the frontier profile index of the null distribution (notebook 6) and, hence, leading to a
conservative FDR threshold. Horizontal lines indicate different values of this FDR threshold.
}
\end{figure}

\begin{enumerate}
    \item Estimate the frontier matrix $F_{ij}$.
    \item Generate a random matrix $R_{ij}$ by shuffling $F$ for pairs of loci
    separated by a genomic distance $s$.
\end{enumerate}

Note that this method over-estimates the frontier profile index of the null
distribution (Fig.~\ref{fig:FI_FDR}, notebook 6), which yields a
conservative FDR threshold.

From the null distribution, we can then estimate for each candidate locus a
p-value and apply multiple testing correction with Benjamini-Hochberg. We can then
filter the set of significant candidate to return only peaks.

\section{Notes}
\begin{enumerate}
\item \label{matrices}
{\it Using matrices.} Using matrices is recommended for several reasons. First, many mathematical
operations, which may be time-consuming using ``for loops,'' are optimized
using matrix operations. This is for instance particular convenient in the
situation of null models where one has to generate and analyze many data to
properly define statistical significance thresholds. Second, matrices often
lead to more compact code and, hence, are more pleasant to work with. We note,
nevertheless, that the size of matrices is limited by the available RAM\@.
Indeed, for $N$ loci, the corresponding Hi-C matrix has a size $N^2$. So, if
one uses $64$ bits-encoding (i.e.~$8$ bytes-encoding) entries, the matrix
occupies $8\times N^2$ bytes. As a consequence, on a regular computer with
e.g. $8$ Giga bytes, the maximum size of the matrix that can be tackled in
principle is $N_{max}=10^{4.5}\simeq30000$ (a $\SI{1}{kb}$ resolution of
\pseudo~Hi-C matrix leads to $N=6276$).

\item \label{colormap} {\it The danger of perceptually non-uniform colormaps.}
One must choose their colormaps with care, as ``colormap are interfaces
between your data and your brain:'' the choices of colors can have a
significant effects on the conclusion you draw from a plot. The colormap
``viridis'' was specifically developed to aid in making well-grounded
decisions from visualization. Specifically, it is a ``perceptually uniform''
colormap, which means that the derivative of the colormap in perceptual space
with respect to the data is uniform. See
\url{https://bids.github.io/colormap/} for more information.

\item \label{outliers}
{\it Identifying Hi-C outliers.} The expected smoothness of $P(s)$ can be used to identify artifacts
coming from the mapping of Hi-C reads. Outliers may indeed lead to large
standard errors of the mean (Fig.~\ref{fig:Ps}). Such mapping problems 
can be simply solved by replacing the problematic
interaction frequency by a typical value at the same genomic distance
(notebook 3.2).

\item \label{diagonals}
{\it Diagonals go by pair for a bacterial chromosome.}
A diagonal index is a positive integer smaller than
$\text{int}(Lg/2/\rho)$. Due to the circularity of chromosome (i.e.\ to
periodic conditions), each $i^{th}$ diagonal goes along with the $(N-i)^{th}$
diagonal, where $N$ is the number of loci.

\item \label{frontiers}
{\it Frontier scales.} Because there exist complex nested sets of interaction domains, frontiers may
actually extend to embed several CIDs, raising {\it en passant} the question
of the (stochastic) nature of interaction domains~\cite{Junier:2015ha}.

\item \label{gradient}
{\it Gradient and periodic conditions of the matrix.}
The \texttt{numpy.gradient()} function does not allow dealing with periodic conditions of the input matrix (here reflecting the circularity of the chromosome). This can nevertheless be handled by defining and using its own gradient function (function \texttt{compute\_gradient()} in notebook 4).

\item \label{fullmulti}
{\it Going beyond the identification of frontiers.}
The current method does not include an analysis of the scale that is associated with a frontier. This analysis is necessary to assess, for instance, whether the frontier corresponds to a unique CID, or wether it corresponds to multiple nested CIDs.

\end{enumerate}

\bibliographystyle{spbasic_unsort}
\bibliography{biblio}

\end{document}